\documentclass{mn2e}

\usepackage{amsfonts}
\usepackage{amsmath}
\usepackage{graphicx}

\newcommand{\p}{\ensuremath{\partial}}
\newcommand{\del}{\ensuremath{\delta}}
\newcommand{\avg}[1]{\ensuremath{\langle \,#1\, \rangle}}
\newcommand{\etal}{\emph{et al}.}
\newcommand{\delc}{\ensuremath{\delta_{\rm c}}}
\newcommand{\gph}{\ensuremath{\Gamma_{\rm PH}}}

\newcommand{\erf}[1]{\ensuremath{{\rm erf}\left(#1\right)}}

\newcommand{\eqn}[1]{equation~\eqref{#1}}

\newcommand{\ph}[1]{\phantom{#1}}

\newcommand{\be}{\begin{equation}}
\newcommand{\ee}{\end{equation}}

\title[Bias with correlated steps]
      {Halo bias in the excursion set approach with correlated steps}

\author[A. Paranjape \& R. K. Sheth]
{Aseem Paranjape$^{1}$\thanks{E-mail: aparanja@ictp.it} 
 \& Ravi K. Sheth$^{1,2}$\\
 $^1$ The Abdus Salam International Center for Theoretical Physics, 
      Strada Costiera 11, 34151 Trieste, Italy\\
 $^2$ Center for Particle Cosmology, University of Pennsylvania, 
      209 S. 33rd St., Philadelphia, PA 19104, USA}
\begin{document}
\pagerange{\pageref{firstpage}--\pageref{lastpage}}

\maketitle 

\label{firstpage}

\begin{abstract}
In the Excursion Set approach, halo abundances and clustering are 
closely related.  This relation is exploited in many modern methods 
which seek to constrain cosmological parameters on the basis of the 
observed spatial distribution of clusters.  However, to obtain 
analytic expressions for these quantities, most Excursion Set based 
predictions ignore the fact that, although different $k$-modes in the 
initial Gaussian field are uncorrelated, this is not true in real space:  
the values of the density field at a given spatial position, when 
smoothed on different real-space scales, are correlated in a nontrivial 
way.  We show that when the excursion set approach is extended to 
include such correlations, then one must be careful to account for 
the fact that the associated prediction for halo bias is explicitly a 
real-space quantity.  Therefore, care must be taken when comparing 
the predictions of this approach with measurements in simulations, 
which are typically made in Fourier-space.  We show how to correct 
for this effect, and demonstrate that ignorance of this effect in 
recent analyses of halo bias has led to incorrect conclusions and 
biased constraints.  
\end{abstract}


\section{Introduction}
The abundance of massive galaxy clusters is expected to be related 
to the abundance of sufficiently overdense isolated regions in the 
initial conditions (Press \& Schechter 1974).  
In the excursion set approach (Epstein 1983; Bond \etal\ 1991; 
Lacey \& Cole 1993), the abundance ${\rm d}n(m)/{\rm d}m$ of such 
overdense `clouds' containing mass $m$ that are not contained within 
larger overdense clouds is estimated by mapping the problem to one 
involving random walks.  If $f(s){\rm d}s$ denotes the first 
crossing distribution of a barrier of a certain prescribed height by 
random walks, i.e., the fraction of walks which have crossed a barrier
$B(s)$ in the `time' interval $(s,s+{\rm d}s)$, without having crossed
it before, then the Excursion Set ansatz is that
\begin{equation}
 \frac{m}{\bar\rho}\frac{{\rm d}n(m)}{{\rm d}m} {\rm d}m = f(s)\,{\rm d}s, 
\end{equation} 
where $\bar\rho$ is the background density.  
In this approach, the barrier height and its dependence on $s$ is 
set by the competition between gravitational collapse and the 
cosmological expansion history (Sheth, Mo \& Tormen 2001), and the 
relation between random walk time $s$ and halo mass $m$ is set by the 
statistics of the initial fluctuation field.  

To obtain analytic expressions for the first crossing distribution, 
Bond \etal\ (1991), and most analyses which followed, assume that 
the density field smoothed on one scale is trivially correlated with
that on another scale.  Namely, if one plots the overdensity $\delta$ 
as a function of smoothing scale $s$, then this resembles a random 
walk in `time' $s$ -- the usual assumption is that successive steps 
in the walk are independent of the previous ones.  This assumption is 
only correct for a Gaussian field smoothed with a series of filters 
that are sharp in $k$-space; it is incorrect for any other, more 
physically reasonable, set of smoothing filters such as a Gaussian 
or a TopHat in real-space (Bond \etal\ 1991).  
For walks with uncorrelated steps, the first crossing problem can be 
solved exactly for barriers whose height is a constant (Bond \etal\ 1991) 
or a linear (Sheth 1998) function of random walk time.  
In particular, the first crossing distribution of 
 $B(s) = \delc\,(1 + \beta s/\delc^2)$, 
for some slope $\beta$, is 
\begin{equation}
 sf_{u}(s) = \frac{B(0)}{\sqrt{2\pi s}}\,{\rm e}^{-B^2(s)/2s}\,,
\label{fudef}
\end{equation}
where the subscript $u$ is to remind us that this is for walks with 
uncorrelated steps.  Reasonably accurate analytic approximations to 
the exact solution for more general barriers are also available 
(Sheth \& Tormen 2002; Lam \& Sheth 2009).  

In this approach, the question of halo bias reduces to the problem of 
estimating the first crossing distribution when the walk starts from 
some prescribed height on some prescribed scale $(\delta,S)$, 
rather than from the origin (Mo \& White 1996; Sheth \& Tormen 1999).  
The halo bias factors $b_k$ are defined by expanding the conditional 
distribution in a Taylor series around $\delta=0$ and $S=0$ and 
ignoring all terms of order $S/s$:  
\begin{equation}
 \sum_{k>0} \frac{b_k}{k!}\,\delta^k \equiv  \frac{f(s|\delta,S)}{f(s)} - 1.
 \label{bkbetter}
\end{equation}
The coefficient of the first term in this series, the linear bias 
factor $b_1$, is the one of most interest.  

For a barrier of constant height $B(s)=\delc$ and walks with 
uncorrelated steps, except for a shift of origin, the conditional 
walk follows the same statistics as the unconditioned one.  
Namely, one simply sets $\delc\to \delc-\delta$ and $s\to s-S$ in 
the expression for the first crossing distribution.  
Therefore, in the $S\ll s$ and $\delta\ll \delc$ limits, 
instead of expanding the conditional distribution in a Taylor series 
(as in equation~\ref{bkbetter}), 
the halo bias factors may equally well have been obtained by 
differentiating the unconditional distribution with respect to 
$\delc$.  This remains true for more general barriers:  
if we use $\delc$ to denote the barrier height on very 
large scales, i.e., $\delc=B(0)$, then 
\begin{equation}
 b_u = -\frac{\partial \ln f_u(s)}{\partial \delc} 
     = \frac1{\delc}\left(\frac{\delc^2}{s} + \beta - 1\right),
 \label{bu}
\end{equation}
where the derivative with respect to $\delc$ actually means 
a derivative with respect to $B(0)$, and the final expression on the 
right assumes $B(s) = \delc\,(1 + \beta s/\delc^2)$.  
(When $\beta=0$, then our $b_u$ reduces correctly to the bias formula 
given by Cole \& Kaiser 1989 and Mo \& White 1996 for a constant barrier.)
It is in this sense that a model for halo abundances carries with it 
a model for halo bias.  
And it is this close relation between halo abundance and bias which 
vastly simplifies analyses which seek to constrain cosmological 
parameters from cluster abundances and clustering.

Recent work has shown that this close relationship 
is not accurate at ten percent precision (Manera \etal\ 2010).  
This is one of the reasons why there is renewed interest in going 
beyond the assumption of uncorrelated steps.  
Peacock \& Heavens (1990) provided an analytic approximation to the 
first crossing distribution of a barrier of constant height, when the 
steps are correlated because of a Gaussian smoothing filter.
Recently, Maggiore \& Riotto (2010) have provided another approximate 
solution to this problem, which is better motivated for a real-space 
TopHat.  
As Paranjape \etal\ (2011) discuss, the two approaches may be thought 
of as expanding around the two opposite limits of completely correlated 
and completely stochastic walks, respectively.  Since Monte Carlo 
solutions of the exact answer show that the first crossing distribution 
for a TopHat is always within about ten percent that for a Gaussian 
(e.g. Bond \etal\ 1991), the question arises as to which approach is 
the more accurate.  Paranjape \etal\ showed that it is the older 
Peacock-Heavens approximation which is the more accurate.  

Recently, Ma \etal\ (2011) have extended the Maggiore-Riotto approach 
to derive expressions for halo bias.  They find that this extension 
appears to predict larger bias factors than for the case of 
uncorrelated steps (i.e., smoothing filters that are sharp in $k$-space).  
Since the Peacock-Heavens approximation to the first crossing 
distribution is the more accurate, the main goal of the present paper 
is to compute the associated prediction for halo bias, to see if it 
too leads to enhanced bias factors.  We do this in Section~\ref{bias}, 
finding that it does.

However, our analysis highlights an important property of smoothing 
with a sharp-$k$ filter.  Namely, for this filter, halo bias does not 
depend on whether one measures it in real-space, by measuring the ratio 
of halo counts-in-cells conditioned to have a certain overdensity to the 
unconditioned average, or from the ratio of the halo-halo or halo-mass 
power spectrum to that of the mass.  
The excursion set prediction is explicitly for the former, whereas 
measurements in simulations are almost always of the latter.  We show 
that when one uses anything other than a sharp $k$-space filter then 
the bias measured from halo counts-in-cells differs from that measured 
using power spectra.  This is the physical origin of the enhanced 
bias associated with TopHat or Gaussian smoothing filters.  
A final section discusses some implications.  

\section{Halo bias and constrained walks}\label{bias}

For what follows, it will be useful to define 
\begin{equation}
 \sigma_j^2(R) \equiv 
 \int \frac{{\rm d}k}{k}\,\frac{k^3P_{\delta}(k)}{2\pi^2}\,k^{2j}\,W^2(kR),
\end{equation}
where $P_{\delta}(k)$ is the power spectrum of initial density
fluctuations (linearly extrapolated to present epoch), and $W$ is a
smoothing filter.  The quantity $\sigma_0^2(R)$ measures 
the variance in the field on scale $R$.  We will often denote this as 
$s=\sigma_0^2$, and will only show the $R$-dependence when omitting 
it would have led to ambiguity. 
In addition, we will also make frequent use of the correlation scale 
defined by Peacock \& Heavens (1990):
\begin{equation}
 \gph \equiv 2\pi\ln(2)\,\sqrt{\gamma^2/(1-\gamma^2)}\,,
 \label{Gph}
\end{equation}
where 
\begin{equation}
 \gamma \equiv \frac{\sigma_1^2}{\sigma_0\sigma_2}\,.
\end{equation}
For Gaussian smoothing filters and $P_{\delta}(k)\propto k^n$, which we 
will use to illustrate our results, $\sigma_j^2\propto R^{-3-n-2j}$ and 
 $\gamma^2 = (3+n)/(5+n)$, making
 $\gph = 2\,\pi\ln(2)\,\sqrt{(n+3)/2}$.

\subsection{Constrained walks with correlated steps}
The analysis below is restricted to barriers $B(s)$ for which 
$B(s)/\sqrt{s}$ decreases monotonically.   Define 
\begin{equation}
 P(s|\delta,S) \equiv \frac12\left[1 + \erf{\frac{\nu_{10}}{\sqrt{2}}}\right],
 \label{Psurvive}
\end{equation}
where
\begin{equation}
 \nu_{10} = \frac{B(s) - r\, \delta\,\sqrt{s/S}}{\sqrt{s(1 - r^2)}}\,,
 \label{v10}
\end{equation}
and 
\begin{equation}
 r = (sS)^{-1/2}\,
     \int \frac{{\rm d}k}{k}\,\frac{k^3\, P_{\delta}(k)}{2\pi^2}\,
     W(kR_s)W(kR_S)\,. 
\end{equation}
Let $f(s|\delta,S)$ be the first crossing distribution of the barrier
$B(s)$ for walks with correlated steps that are constrained to pass
through some $(\delta,S)$ with $S<s$, while also remaining below the
barrier $B(S^\prime)$ on scales $S^\prime<S$. Then a good
approximation to $f(s|\delta,S)$ is,
\begin{equation}
 f(s|\delta,S) = 
 -\frac{\partial}{\partial s} \bigg(P(s|\delta,S)\,
          E(s|\delta,S)\bigg)\,,
 \label{fph}
\end{equation}
where 
\begin{equation}
 E(s|\delta,S) \equiv 
   \exp\left(\int_0^s \frac{{\rm d}s^\prime}{s^\prime\gph}\,
             \ln P(s^\prime|\delta,S)\right)\,,
 \label{Ecorrection}
\end{equation} 
and $\gph$ was defined in equation~(\ref{Gph}) (Paranjape
\etal\ 2011). 

The unconditional first crossing distribution $f(s)$ is given by
setting $r=0$ in the expression for $\nu_{10}$:
\begin{align}
 P(s) &= \frac12\left[1 +
   \erf{B(s)/\sqrt{2s}}\right] \,, \label{Pbuc}\\ 
 E(s) &= \exp\left(\int_0^s \frac{{\rm
     d}s^\prime}{s^\prime\gph}\, \ln P(s^\prime)\right)
 \,,\label{Ebuc}\\ 
sf(s) &= -s\frac{\p}{\p s}\left(P(s)\,E(s)\right)\nonumber\\
       &=E(s)\left[sf_{\rm c}(s) - \frac1{\gph}P(s)\ln
  P(s)\right]\,,\label{fbuc} 
\end{align}
where
\be
 sf_{\rm c}(s)% &=& 
=-\frac{\p P}{\p \ln s}
            = -\frac{\partial\ln [B(s)/\sqrt{s}]}{\partial\ln s}\,
           \frac{B(s)}{B(0)}\, sf_{u}(s) \,.
\label{fcdef}
\ee
The limit $\gph\to \infty$ is what Paranjape \etal\ (2011) 
called completely correlated walks.  In this limit  $E\to 1$: 
the quantity $f_{\rm c}(s)$ is therefore the first crossing distribution 
for completely correlated walks.  
Paranjape \etal\ showed that this limit provides an excellent
description of the unconditional first crossing distribution of barriers 
which decrease steeply with $s$. In particular, for such barriers, this 
limit is a better description of $f(s)$ than if one uses the value 
for $\gph$ that is given by equation~(\ref{Gph}).  
This will be important below. 

Note that $f_{\rm c}\ne f_{u}$.  For barriers of the form
\be
 B(s)=\delc\left(1+\beta(s/\delc^2)^\alpha\right)\,,
\label{alphabarrier}
\ee
where we assume $\alpha > 0$ and $\beta\leq 0$, 
\begin{align}
 sf_{\rm c}(s) &=\frac{sf_u(s)}{2}
         \left(1-\frac{\delc}{B(0)}\beta(2\alpha-1) (s/\delc^2)^\alpha \right)
         \nonumber\\
      &=\frac{sf_u(s)}{2} \left(1-\beta(2\alpha-1) (s/\delc^2)^\alpha\right) \,,
\label{fcalphabarrier}
\end{align}
where we have explicitly shown the dependence on $B(0)$ in the first
line. For a barrier of constant height ($\beta=0$), 
we have $f_{\rm c} = f_{u}/2$. 
For a linear barrier ($\alpha=1$),
 $f_{\rm c} = (1 - \beta s/\delc^2)\,f_{u}/2$.

\subsection{Bias as the large scale limit}
Notice that the integral in our expression for $r$ is similar to that 
which defines $s = \sigma_0^2(R_s)$ and $S=\sigma_0^2(R_S)$, the only 
difference being that here the two smoothing filters have different 
scales.  If we use $S_\times$ to denote the value of this integral, 
then $r\sqrt{s/S} = S_\times/S$.  
This quantity depends on the form of the smoothing filter.  
For a sharp-$k$ filter (the one associated with uncorrelated steps) 
$S_\times = S$, so 
$\nu_{10} = [B(s) - \delta_0]/\sqrt{s-S}$.  Thus, $\nu_{10}$ is 
related to the unconstrained quantity ($r=0$ in equation~\ref{v10}) 
by a shift of origin.  Therefore, the first term $b_1$ in the expansion 
of equation~(\ref{bkbetter}) will be the same as $b_u$ of 
equation~(\ref{bu}).  

For Gaussian filtering of a power-law spectrum 
   $r = [2R_SR_s/(R_S^2+R_s^2)]^{(n+3)/2}$ 
and $S_\times/S = [2/(1+R_s^2/R_S^2)]^{(n+3)/2}$.  
In this case, if one wishes to think of $\nu_{10}$ as an effective 
shift of origin, then this shift is scale dependent.  However, 
if $R_s\ll R_S$, then $S_\times/S\to 2^{(n+3)/2}$:  in this limit, the 
Gaussian case is similar to the $k$-space one, except that the shift 
of origin is $\delc - 2^{(n+3)/2}\delta$.  This shift comes from the 
fact that steps are correlated.  
As a result, following equation~(\ref{bkbetter}) above will amount 
to the same as differentiating the first crossing distribution with 
respect to $\delc$ (or more generally, with respect to $B(0)$), 
only taking care to account for the factor of $S_\times/S \to 2^{(n+3)/2}$.  

The discussion above applies for other smoothing filters too, 
except that the numerical coefficient $2^{(n+3)/2}$ will change.  
We will argue shortly that the actual value matters little for 
our purposes, but to complete the discussion, we provide some 
explicit examples of this change.  
For a filter that is a tophat in real space, 
$S_\times/S \to 5/4, 4/3$ and 1 for $n=-2, -1$ and 0.
Thus, in general, the numerical coefficient is smaller than when 
the filter is Gaussian.  But the important point is that, for walks 
with correlated steps, equation~(\ref{bu}) should be replaced by 
\begin{equation}
 b_1 = -\left(\frac{S_\times}{S}\right)\,
        \frac{\partial \ln f(s)}{\partial \delc},
 \label{b1}
\end{equation} 
where $S_\times/S \sim 1$ depends on the filter and the power
spectrum, and $f(s)$ is the unconditional first crossing
distribution. Note that, to obtain this result, we use \eqn{Psurvive}
in the limit $S\ll s$, so that we have $\nu_{10}\to(\delc-\del
S_\times/S)/\sqrt{s}$ where $S_\times/S$ is constant. Plugging this
into \eqn{fph}, expanding to linear order in \del, and using
\eqn{bkbetter} finally results in \eqn{b1}.

\subsection{Bias in real vs Fourier space}
To better appreciate the origin of the extra $S_\times/S$ factor,
it is useful to consider the case of peaks in the initial Gaussian
fluctuation field.  Since the property of being a peak is
scale-dependent, we must first specify the smoothing-scale and
filter with which the peak was identified.  We will use $R_{\rm pk}$
to denote this scale.

Bardeen \etal\ (1986; hereafter BBKS) provide expressions for the
mean density profile around a peak (their equation 7.8 and Appendix D) --
as they note, this is essentially the same as the cross-correlation
function between peaks and the dark matter field -- and for how the
large scale density modulates the abundance of peaks (their
Appendix E).  Each of these is explicitly a real-space statement,
and so BBKS provide different expressions for each case.  Note that
the former is one way in which peak-bias is defined, and the latter
is close in spirit to the conditional crossing calculation in the
previous section, where we were interested in how the first crossing
distribution was modulated by knowing the value of the density field
on some large smoothing scale.

However, Desjacques \etal\ (2010) show that all these BBKS expressions
are in fact a consequence of the fact that peaks may be thought of as
being linearly biased tracers of the dark matter, with a $k$-dependent
bias factor:
\begin{equation}
 \delta_{\rm pk}({\bf k}) = b_{\rm pk}({\bf k})\,\delta_m({\bf
k})\,W(kR_{\rm pk})\,.
 \label{peakbias}
\end{equation}
The property of being a peak is scale-dependent, which is why the
smoothing scale appears explicitly in the expression for the
peak field.  And note that the value of the bias factor actually
depends on properties of the peak (height and curvature), so it
depends implicitly on $R_{\rm pk}$.
That is to say, peak-bias is simple in Fourier space,
and all the key BBKS expressions for real-space quantities (e.g.
in their Appendices D and E) are simply consequences of multiplying
the above bias relation with different filter functions before
Fourier transforming to obtain the real-space expressions.

In particular, the cross-correlation between peaks (defined on
scale $R_{\rm pk}$) and the mass smoothed on scale $R_S$ is
\begin{equation}
 \Bigl\langle\delta_{\rm pk}\delta_S\Bigr\rangle =
  \int \frac{{\rm d}k}{k}\,b(k)\,\frac{k^3\, P_{\delta}(k)}{2\pi^2}\,
        W(kR_{\rm pk})W(kR_S)
\end{equation}
where we have defined $\delta_S\equiv\delta_m(R_S)$.
In the high-peak limit (and in the $k\ll 1$ limit), $b(k)$ becomes
independent of $k$.  In this case
\begin{equation}
 \Bigl\langle\delta_{\rm pk}\delta_S\Bigr\rangle \to b\, S_\times
\end{equation}
and so
\begin{equation}
 \Bigl\langle\delta_{\rm pk}|\delta_{S}\Bigr\rangle
  \to \frac{\langle\delta_{\rm pk}\delta_{S}\rangle}
         {\langle\delta_{S}^2\rangle}\,\delta_S
  \to b\,\frac{S_\times}{S}\,\delta_S.
\label{peaks-lim}
\end{equation}
This shows that when the Fourier space bias factor is $b$, then in
real space, the bias picks up a factor of $S_\times/S$.

It is usual to assume that high peaks are not a bad model for massive
halos.  So it is not unreasonable to assume that the expression above,
with the replacement $\delta_{\rm pk}\to \delta_h$ is appropriate for
halo bias. Since the excursion set approach identifies $\del_h$ with
the right hand side of \eqn{bkbetter}, comparison of \eqn{peaks-lim}
with equation~(\ref{b1}) suggests that $\partial \ln
f(s)/\partial\delc$ returns the more fundamental Fourier space bias
factor, even though $f(s)$ was estimated from a real-space
calculation; the factor $S_\times/S$ arises simply because the
excursion set calculation returns the cross-correlation between halos
(regions in the initial conditions which are above $\delta_{\rm c}$ on
scale $R_s$) and the initial mass fluctuation field (smoothed on
$R_S\gg R_s$). 

In the Appendix, we discuss the
consequences of \eqn{peakbias} with $\del_{\rm pk}\to\del_h$ and
$R_{\rm pk}\to R_h$, for real-space $2$-point correlation
functions (in which the matter field is \emph{not} smoothed on some
large scale $R_S$). In particular, the bias determined from such
real-space correlation functions will depend on halo mass, with the
differences between the different measurements being in qualitative 
agreement with results of $N$-body simulations (Manera \etal\ 2010).

\subsection{Halo bias for constant or decreasing barriers}
Having motivated why $\partial \ln f(s)/\partial\delc$ is fundamental 
even for walks with correlated steps, we now evaluate it for a few 
special cases. (Recall that the derivative here is actually with 
respect to $B(0)$). The general expression is 
\begin{align}
 \frac{\partial \ln f(s)}{\partial \delc} &=
\frac{\partial \ln E(s)}{\partial \delc} \nonumber\\
&\ph{100}
+ \left(\frac{\partial\ln f_{\rm c}(s)}{\partial\delc} -
  \frac1{\gph}\frac{\partial P}{\partial\delc}\frac{(1+\ln
   P)}{sf_{\rm c}(s)}\right)  \nonumber\\
&\ph{100\frac{\partial \ln E(s)}{\partial \delc}}\times 
\left(1 - \frac1{\gph}\frac{P\,\ln P}{sf_{\rm c}(s)} \right)^{-1}\,,
 \label{bph}
\end{align}
where $P(s)$ and $f_{\rm c}(s)$ were defined in equations~(\ref{Pbuc})
and \eqref{fcdef}, respectively.

For the barrier of \eqn{alphabarrier}, we have 
\begin{align}
 \frac{\partial \ln E(s)}{\partial \delc} 
     &=\frac1{\delc\gph}\int_0^s {\rm d}s^\prime\,
         \frac{f_{u}(s^\prime)}{P(s^\prime)}\,,\nonumber\\
 \frac{\partial P(s)}{\partial\delc} 
   &= \frac{1}{\delc}sf_u(s)\,,\nonumber\\
 \frac{\partial\,\ln f_{\rm c}(s)}{\partial\delc}
&=\frac1{\delc}\left(\frac{sf_{u}(s)}{2sf_{\rm c}(s)}\right)\nonumber\\
&\ph{1}\times \bigg(1-\delc^2/s + 2\beta(\alpha-1)
   (s/\delc^2)^{\alpha-1}\nonumber\\
&\ph{1-(\delc^2/s)}
 + \beta^2(2\alpha-1)(s/\delc^2)^{2\alpha-1}\bigg)\,.  
\label{ddcdetailslinear}
\end{align}
For the constant barrier ($\beta = 0$), $\partial\ln
E(s)/\partial\ln\delc = -2\,\ln P(s)/\gph$.
In general, equation~(\ref{bph}) shows that, 
as $\gph\to\infty$, $\partial \ln f(s)/\partial \delc\to 
  \partial \ln f_{\rm c}(s)/\partial \delc$, as it should 
(the term involving the derivative of $E$ is always proportional to
$1/\gph$).  For linear barriers ($\alpha=1$), this limit provides 
a better description of the first crossing distribution for sufficiently 
negative $\beta$ (Paranjape \etal\ 2011).  Therefore, we might expect 
this to be true for the bias factor also.  In this case,
\be
 \frac{\partial\,\ln f_{\rm c}(s)}{\partial\delc}
   =\frac1{\delc}\left(\frac{1 - \delc^2/s + \beta^2
         s/\delc^2}{1-\beta s/\delc^2}\right)\,. 
\label{ccbias}
\ee

\begin{figure*}
 \centering
 \includegraphics[width=0.46\vsize]{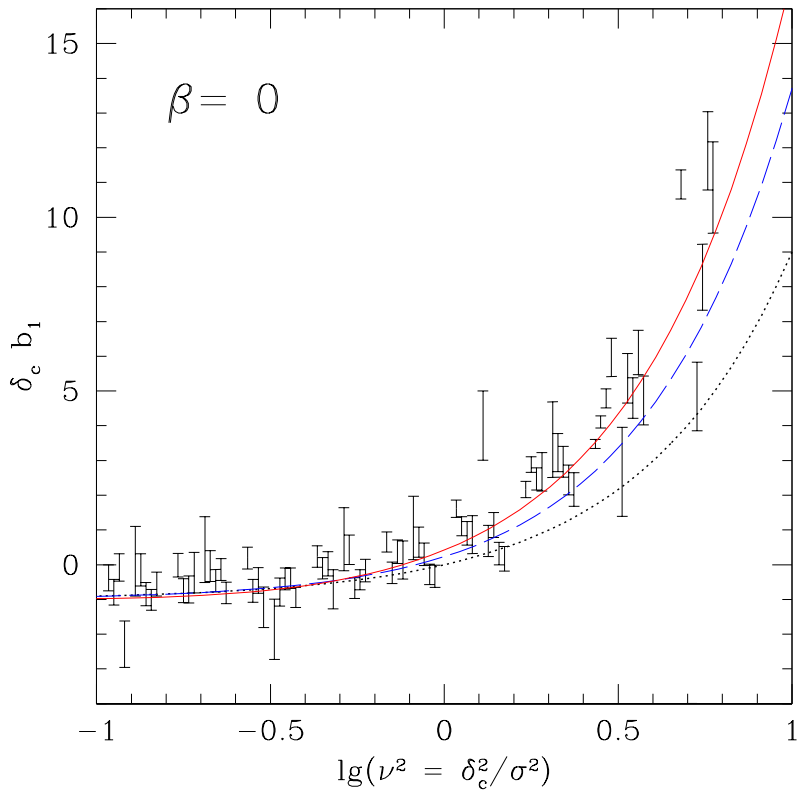}
 \caption{Large-scale linear bias factor associated with a barrier 
          of constant height $\delc$.  
          Symbols with error bars show the bias factor determined via 
          Monte-Carlo solution of equation~(\ref{bkbetter}), for walks 
          with correlated steps because of Gaussian smoothing and 
          $P_{\delta}(k)\propto k^{-1.2}$.  
          I.e., $\delc b_1$ was determined by considering walks 
          conditioned to pass through $\delta\ll \delc$ on 
          scale $S\ll \delc^2$, for a range of choices of $\delta$.  
          Specifically, we show
           $[f(s|\delta,S)/f(s) - 1]/(\delta/\delc)$
          for $S = (0.15 \delc)^2$ and $\delta/\delc$ 
          bin centers ranging from $[-1.8,-0.2]\sqrt{S}$, and from 
          $[0.2,1.8]\sqrt{S}$, in steps of $0.4\sqrt{S}$, each of 
          width $0.2\sqrt{S}$.
          (For clarity, at each $\nu$, 
          symbols for different $\delta$ have been offset slightly.)
          Dotted curve shows $\delc b_u = \nu^2-1$, which would 
          be appropriate for walks with uncorrelated steps 
          (equation~\ref{bu}).  
          Solid curve shows the result of inserting equation~(\ref{bph}) 
          with $\gph = 4.3$ in equation~(\ref{b1}).
          Dashed curve shows $\delc b_1$ determined from 
          equation~(24) of Ma \etal\ (2011) with their $\kappa = 0.35$. }
 \label{b1ph}
\end{figure*}

\subsection{Comparison with Monte-Carlo solution}
To test our analysis, Figure~\ref{b1ph} compares the bias factor 
we measure by direct Monte-Carlo simulation of $b_1$ via 
equation~(\ref{bkbetter}).  I.e., we construct the first crossing 
distribution of $B(s) = \delc$ by walks conditioned to pass 
through some $\delta\ll \delc$ on scale $S\ll \delc^2$.  
Then we divide this by the (Monte-Carlo estimate of the) unconditional 
first crossing distribution, and subtract 1.  Our Monte-Carlos assumed 
a power spectrum of the form
 $P_{\delta}(k)\propto k^{-1.2}$ and Gaussian smoothing filters, for which 
 $S_\times/S \to 2^{0.9}$ and $\gph=4.3$.  
The solid curve shows the result of inserting equation~(\ref{bph}) 
with $\gph=4.3$ in equation~(\ref{b1}).  It provides an excellent 
description of our measurements.   
The dotted curve shows the bias relation for uncorrelated walks 
(i.e. a sharp-$k$ filter), equation~(\ref{bu}), is noticably different, 
lying well below the measurements at high $\delc/\sigma$.  

The dashed curve shows the corresponding expression for the bias 
from Ma \etal\ (2011): the right hand side of their equation~(24) 
minus 1, times $\delc$.   Their expression has one free parameter, 
$\kappa$; we set it to 0.35, the value they say is appropriate for 
Gaussian smoothing filters.  
Notice that, although it is not as good a description of our 
measurements as is our solid curve, it does a reasonable job, 
indicating that the technical machinery which went into generating 
this prediction has produced a reasonably accurate result.  
However, this same technical machinery has obscured the reason why 
their formula predicts a bias factor that is approximately twice as 
large as that given by equation~(\ref{bu}) when $\delc/\sigma \gg 1$.  
Ma \etal\ attribute this difference to what they call stochasticity 
in the barrier height -- but this cannot be correct, because 
stochasticity is {\em not} present in our constant barrier Monte-Carlos.  

To make this point more directly, note that on the large scales which 
are appropriate for the bias calculation, our analysis shows that the
conditional distribution $f(s|\delta,S\to 0)$ should be very well
approximated by the unconditional distribution $f(s)$ with a shift of
origin by $(S_\times/S)\, \delta$.  Upon noting that Ma et al.'s $\kappa$
is our $S_\times/S - 1$, it is easy to show that their equation~(A32)
for $f(s|\delta,S\to 0)$, is indeed what our analysis suggests, except
that they have used their own expression for $f(s)$, and the result
has been expanded to lowest order in $\kappa$.  However, this 
expansion in $\kappa$ has obscured the simple, intuitive relation 
between the conditional and unconditional distributions that our 
analysis exploits (e.g. Ma \etal\ themselves failed to notice it).  
This also explains why their expression for the bias factor does not 
provide as good a description of our Monte-Carlos as does ours:
their expression assumes $\kappa\ll 1$ when it is not (this matters
more than the fact that their expression for $f(s)$ is not as accurate
as ours).

\begin{figure}
 \centering
 \includegraphics[width=0.9\hsize]{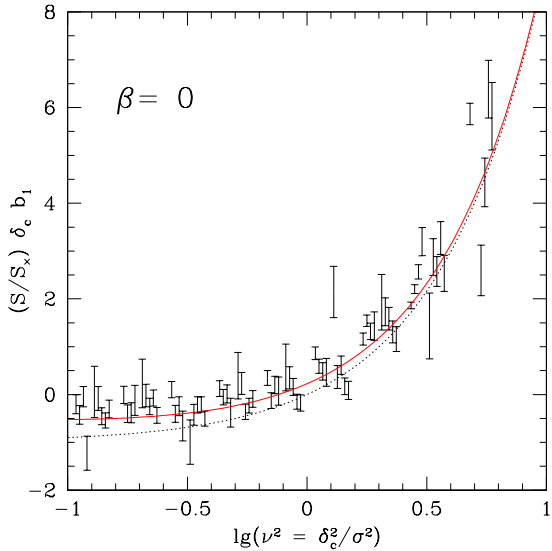}
 \caption{Same as previous Figure, except that now we have rescaled 
          all quantities by a factor of $S/S_\times$.  For the measurements 
          and the solid curve, this factor is $1/2^{0.9}$; for the dotted 
          curve it is $1$.  }
 \label{biasph}
\end{figure}

The results of the previous section suggest that the main difference 
between the solid and dotted curves (for Gaussian and sharp-$k$ filtered 
walks) is almost entirely due to the fact that the excursion set 
measurement yields a real-space bias factor which carries an 
additional factor of $S_\times/S$:  for our Gaussian smoothed walks, 
this factor is $2^{0.9}$, whereas it is unity for the sharp-$k$ filter 
associated with equation~(\ref{bu}).  
To show this explicitly, Figure~\ref{biasph} shows the result of 
dividing our measurement by $S_\times/S$ so as to compare it with 
the derivative of the unconstrained first crossing distribution, 
which we compute analytically using equation~(\ref{bph}), and show 
as a solid curve.  Notice that now the measurements (and our solid 
curve) are much closer to the dotted curve (which shows 
equation~\ref{bu}).  The differences between the solid and dotted 
curves in this plot are entirely a consequence of the fact that the 
underlying first crossing distributions are different;  they are a 
much fairer depiction of the effect of correlated steps on the 
large scale bias.  
The graphic differences between Figures~\ref{b1ph} and~\ref{biasph} 
demonstrate explicitly that correctly accounting for the factor of 
$S_\times/S$ is essential; failure to do so leads to incorrect 
conclusions about the nature of the relationship between the 
derivative of the first crossing distribution and the large scale 
bias.  

\begin{figure}
 \centering
 \includegraphics[width=0.9\hsize]{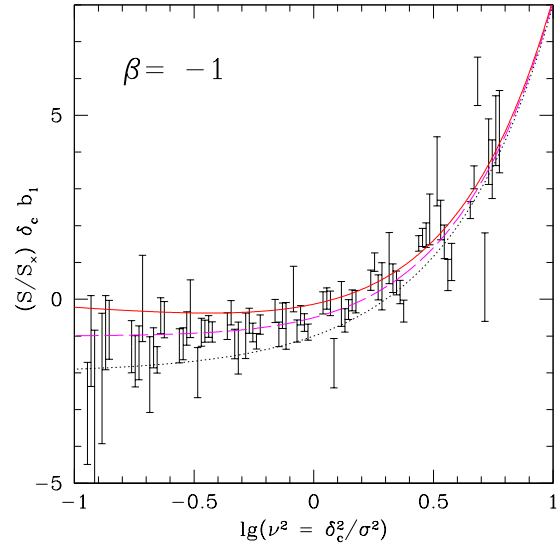}
 \caption{Same as Figure~\ref{biasph} except now the barrier is 
          $B(s) = \delc(1 + \beta s/\delc^2)$.
          Symbols with error bars show the bias factor determined via 
          Monte-Carlo solution of equation~(\ref{bkbetter});
          dotted curve shows the prediction for walks with 
          uncorrelated steps (from equation~\ref{bu});  
          solid curve shows equation~(\ref{bph}), and dashed 
          curve shows $\partial \ln f_{\rm c}(s)/\partial\ln\delc$
          (from \eqn{ccbias}), which is the $\gph\to\infty$ limit of 
          equation~(\ref{bph}).}
 \label{biasLinear}
\end{figure}

Figure~\ref{biasLinear} shows a similar analysis of the bias associated 
with a linearly decreasing barrier (we set $\alpha=1$ and $\beta=-1$).  
In this case, $b_u$ of equation~(\ref{bu}) lies well below the 
Monte-Carlo solution, whereas $b_1$ of equation~(\ref{bph}) lies above 
it, at least at small $\delc/\sigma$.   However, the $\gph\to\infty$ 
limit of equation~(\ref{bph}), namely 
 $\partial\ln f_{\rm c}(s)/\partial\delc$ from \eqn{ccbias}, 
provides an excellent description.  
This is very reassuring because, as we remarked previously, $f_{\rm c}(s)$ 
itself (the $\gph\to\infty$ limit of the first crossing distribution) 
also provides a better description of the Monte-Carloed first crossing 
distribution.  We note again that accounting for the factor of 
$S_\times/S$ was crucial.  

\section{Discussion}
We have presented an analysis of the large scale bias factor associated 
with the Excursion Set approach for smoothing filters which give rise 
to walks with correlated steps.  This required careful consideration 
of the differences between real and Fourier space bias -- something 
that has been all but ignored to date since, for a sharp-$k$ filter, 
these two are the same.  For other filters, the real-space bias will 
generically be different.  

In particular, our analysis shows that the first crossing distribution 
$f(s|\delta,S)$, of walks that are constrained to pass through $\delta$ 
on scale $S$ should, as $S\to 0)$, be very well approximated by the 
unconditional distribution $f(s)$ with a shift of origin by 
$(S_\times/S)\, \delta$ (equations~\ref{Psurvive} and~\ref{v10}).  
Thus, memory effects in the walks manifest in two conceptually 
different ways:  the fact that $S_\times/S\ne 1$, and that the 
functional form of $f(s)$ itself is altered.  We argued that the former 
is simply associated with smoothing effects associated with the precise 
way in which the bias factor was defined; the latter is the more 
fundamental difference.  
The usual procedure of differentiating the mass function with respect 
to the height of the walk on large scales provides a good description 
of the latter, which we identified with the Fourier-space bias factor; 
it underestimates the real-space bias estimated by the cross correlation 
between halos and the mass (smoothed on large scales), by a factor 
$S_\times/S$ that can be as large as 2 (\eqn{peaks-lim}; this factor 
is slightly smaller for $\Lambda$CDM-like power spectra), while 
slightly overestimating the bias from real-space correlation functions 
(\eqn{app-realspacebias}). 

Comparison with measurements in Monte Carlo simulations of 
constrained random walks showed good quantitative agreement with our 
excursion set predictions of the real-space bias.  In particular, 
the effect of the $S_\times/S$ term (the first of our two effects) 
can be large (compare the solid and dotted curves in Figure~\ref{b1ph}).  
The fact that the first crossing distribution $f$ (\eqn{fbuc}) is not 
the same as the one for uncorrelated steps $f_u$ (\eqn{fudef}) also 
matters, though it is subdominant (solid and 
dotted curves in Figure~\ref{biasph} are different, but this difference 
is smaller than in Figure~\ref{b1ph}).  
However, this second effect is the only one which matters for the 
Fourier space bias.  
The agreement between the solid curves and the measurements in 
Figures~\ref{b1ph} and~\ref{biasph} suggest that any additional effects 
missed by our analysis must be small.  

Our results highlight the fact that, when bias is simple in Fourier 
space, then, once one goes beyond sharp-$k$ smoothing, almost all 
measures of the large-scale bias factor will differ from one another, 
but these differences can be understood simply as a result of 
applying different smoothing filters to the same underlying Fourier 
space bias.  Differences between real and Fourier space measures of 
the large scale bias factor are now being seen in simulations
(e.g. Manera \etal\ 2010), so future comparison of these bias factors 
with simulations should be careful to specify which measure of the 
bias is actually being studied.  

We noted that our excursion set calculation of the difference between 
real and Fourier space bias is consistent with the differences between 
real and Fourier space bias for peaks reported by 
Desjacques \etal\ (2010).  The peaks calculation is remarkable 
because there, the bias factors are $k$-dependent.  We believe that 
one of the virtues of our excursion set analysis is that it shows 
clearly how the excursion set approach too can lead to $k$-dependent 
bias.  Namely, if halos of mass $m$ formed from peaks in the initial 
density field, then one would replace equation~(\ref{v10}) with the 
expression which describes the density run around peaks rather than 
random positions of height $\delc/\sigma(m)$.  This would yield 
real-space bias factors which, following Desjacques \etal\ (2010), 
could then be interpreted in terms of $k$-dependent bias in 
Fourier space, in effect providing the first calculation of 
$k$-dependent bias from excursion set theory.  

\section*{Acknowledgments}
We thank T. Y. Lam and Roman Scoccimarro for discussions about walks
with correlated steps, and the referee for a helpful report. 
RKS is supported in part by NSF-AST 0908241,
and thanks B. Bassett and AIMS for their hospitality in early April
2011.

\appendix

\section{Halo-mass cross correlation with Fourier-space biasing}
\noindent
When the halo field in Fourier-space is linearly biased as in
\eqn{peakbias} with $\del_{\rm pk}\to\del_h$ and $R_{\rm pk}\to R_h$,
and with a constant bias $b$, then the bias determined from the
ratio of the halo-mass cross correlation $\xi_{\rm hm}$ and the mass 
autocorrelation function $\xi_{\rm mm}$, will in general be smaller
than $b$. To see this, consider a toy model for the CDM power spectrum
given by $P_{\delta}(k)\propto k\,{\rm e}^{-k^2R_{\rm C}^2/2}$. 
Further assuming the halo field in \eqn{peakbias} to be Gaussian 
filtered with $W(kR_h) = {\rm e}^{-k^2R_h^2/2}$, we have
\begin{align}
\xi_{\rm mm}(r;R_{\rm C}) &\equiv \avg{\del_m(0)\del_m({\bf r})}
\nonumber\\
& = \int\frac{{\rm d}k}{k}\, \frac{k^3P_{\delta}(k)}{2\pi^2}\, j_0(kr)\nonumber\\
&=\xi(r/R_{\rm C})\,, \label{app-ximm}\\
\xi_{hm}(r;R_{\rm C},R_h) &\equiv \avg{\del_h(0)\del_m({\bf r})}
\nonumber\\ 
& = b\int\frac{{\rm d}k}{k}\, \frac{k^3P_{\delta}(k)}{2\pi^2}\,{\rm e}^{-k^2R_h^2/2}
j_0(kr)\nonumber\\ 
&=b\left(1+\frac{R_h^2}{R_{\rm C}^2}\right)^{-2}
\xi\left(r/\sqrt{R_h^2+R_{\rm C}^2}\right)\,,\label{app-xihm}
\end{align}
with the same function $\xi(x)$ appearing in both expressions, but
with different arguments. For $R_h\ll R_{\rm C}$ we can then write
\be
\frac{\xi_{hm}}{\xi_{mm}} \simeq b\left[1 -
  \left(2+\frac12\frac{d\ln\xi}{d\ln x}\right) \frac{R_h^2}{R_{\rm
      C}^2} + \ldots \right] \,,
\label{app-realspacebias}
\ee
where $x=r/R_{\rm C}$. For separations $r$ such that $\xi(x)\propto
x^{-(n+3)}$, this bias from correlation functions therefore picks up a
relative decrement of $|\Delta b|/b \simeq (|n-1|/2)(R_h/R_{\rm
  C})^2$, whose magnitude depends on halo mass, becoming smaller for
halos of smaller mass. This is qualitatively in agreement with the
results of $N$-body simulations discussed by Manera \etal\ (2010).

\label{lastpage}


\begin{thebibliography}{99}
\bibitem{bond91} Bond J.~R., Cole S., Efstathiou G., Kaiser N., 1991, ApJ, 379, 440 
\bibitem{cole89} Cole S., Kaiser N., 1989, MNRAS, 237, 1127
\bibitem{desj10}  Desjacques V., Crocce M., Scoccimarro R., Sheth R.~K.,
  2010, PRD, 82, 103529
\bibitem{eps83} Epstein R.~I., 1983, MNRAS, 205, 207
\bibitem{lacey93} Lacey C., Cole S., 1993, MNRAS, 262, 627 
\bibitem{lam09} Lam T.~Y., Sheth R.~K., 2009, MNRAS, 398, 2143 
\bibitem{ma11} Ma, C.-P., Maggiore M., Riotto A., Zhang J., 2011, MNRAS, 411, 2644
\bibitem{magg10} Maggiore M., Riotto A., 2010, ApJ, 711, 907
\bibitem{manera10} Manera M., Sheth R.~K., Scoccimarro R., 2010, MNRAS, 402, 589
\bibitem{mo96} Mo H.~J., White S.~D.~M., 1996, MNRAS, 282, 347
\bibitem{paran11} Paranjape A., Lam T.~Y., Sheth R. K., 2011,
  arXiv:1105.1990, MNRAS, in press
\bibitem{peac90} Peacock J. A., Heavens A. F., 1990, MNRAS, 243, 133
\bibitem{press74} Press W. H., Schechter P., 1974, ApJ, 187, 425
\bibitem{sheth98} Sheth R.~K., 1998, MNRAS, 300, 1057 
\bibitem{sheth01} Sheth R.~K., Mo H.~J., Tormen G., 2001, MNRAS, 323, 1 
\bibitem{sheth99} Sheth R.~K., Tormen G., 1999, MNRAS, 308, 119

\end{thebibliography}
\end{document}